\documentclass[pra,superscriptaddress,twocolumn]{revtex4-2}

\usepackage{graphicx}%
\usepackage{epsfig}
\usepackage{amsmath,amssymb,amsfonts}
\usepackage{xcolor}%
\newcommand{\ket}[1]{\left | #1 \right\rangle}
\newcommand{\bra}[1]{\left \langle#1 \right |}

\begin{document}

\title{Unmasking the Polygamous Nature of Quantum Nonlocality}

\author{Pawe{\l} Cie\'sli\'nski}

\affiliation{Institute of Theoretical Physics and Astrophysics, University of Gdańsk, 80-308 Gda\'nsk, Poland}

\author{Lukas Knips}
\affiliation{Max Planck Institute for Quantum Optics, 85748 Garching, Germany}
\affiliation{Faculty of Physics, Ludwig Maximilian University, 80799 Munich, Germany}
\affiliation{Munich Center for Quantum Science and Technology, 80799  Munich, Germany}

\author{Mateusz Kowalczyk}

\affiliation{Institute of Theoretical Physics and Astrophysics, University of Gdańsk, 80-308 Gda\'nsk, Poland}

\author{Wies{\l}aw Laskowski}

\affiliation{Institute of Theoretical Physics and Astrophysics, University of Gdańsk, 80-308 Gda\'nsk, Poland}

\author{Tomasz Paterek}

\affiliation{Institute of Theoretical Physics and Astrophysics, University of Gdańsk, 80-308 Gda\'nsk, Poland}

\affiliation{School of Mathematics and Physics, Xiamen University Malaysia, 43900 Sepang,  Malaysia}

\author{Tam\'as V\'ertesi}

\affiliation{MTA Atomki Lend{\"u}let Quantum Correlations Research Group, Institute for Nuclear Research, 4001 Debrecen, Hungary}

\author{Harald Weinfurter}
\affiliation{Institute of Theoretical Physics and Astrophysics, University of Gdańsk, 80-308 Gda\'nsk, Poland}
\affiliation{Max Planck Institute for Quantum Optics, 85748 Garching, Germany}
\affiliation{Faculty of Physics, Ludwig Maximilian University, 80799 Munich, Germany}
\affiliation{Munich Center for Quantum Science and Technology, 80799  Munich, Germany}

\begin{abstract}
Quantum mechanics imposes limits on the statistics of certain observables. Perhaps the most famous example is the uncertainty principle. Similar trade-offs also exist for the simultaneous violation of multiple Bell inequalities. In the simplest case of three observers, it has been shown that if two observers violate a Bell inequality then none of them can violate any Bell inequality with the third observer,
a property called monogamy of Bell violations. Forms of Bell monogamy have been linked to the no-signalling principle and the inability of simultaneous violations of all inequalities is regarded as their fundamental property.
Here we show that the Bell monogamy does not hold universally and that in fact the only monogamous situation exists for only three observers. Consequently, the nature of quantum nonlocality is truly polygamous.
We present a systematic methodology for identifying quantum states, measurements and tight Bell inequalities that do not obey the monogamy principle for any number of more than three observers. The identified polygamous inequalities enable any subset of $(N-1)$ observers to reveal nonlocality, which is also shown experimentally by measuring Bell-type correlations of six-photon Dicke states. Our findings may be exploited for multiparty quantum key distribution as well as simultaneous self-testing of multiple nodes in quantum networks.
\end{abstract}

\maketitle


\section{Introduction}
\label{sec1}

Quantum nonlocality is one of the most intriguing features of quantum theory. Starting from its beginnings and the famous EPR argument~\cite{EPR1935}, through the first works of John Bell~\cite{Bell1964} to various experiments~\cite{Aspect1982b, Weihs1998, Rowe2001, Pan2000, Walther2005, Hensen2015, Giustina2015, Shalm2015, Rosenfeld2017, Rauch2018}, it reveals the impossibility of a local-realistic description of quantum phenomena. Violation of a Bell inequality serves now not only as a fundamental test for the statements about the nature of reality but also finds applications in many areas of modern quantum technologies~\cite{Ekert1991, Mayers2003, Barrett_2005, Pironio2010, Buhrman_2010, Brunner_2014, Supic_2020}. One crucial concept in the study of quantum nonlocality is the monogamy principle, which states that it is impossible to simultaneously violate all $k$-partite ($k < N$) two-setting Bell inequalities among $N$ different parties. 
The early findings by Scarani and Gisin~\cite{SG2001}, and later by Toner and Verstraete~\cite{TVmono2006,Toner2009mono,Toner2007PhD}, 
reveal that a strict monogamy exists between violations of bipartite Bell inequalities, i.e. if two observers violate a Bell inequality then none of them can violate any Bell inequality with the third observer (note that inequalities with more settings can be violated~\cite{Collins2004}).
A number of generalisations to multipartite scenarios showed that more than one subset of parties can violate a multipartite inequality, but never all subsets of observers~\cite{Kurzynski2011, Augusiak2014, Ramanathan2014, Tran2018, Augusiak2017, Ram_2018, Wiesniak_2021}. These forms of monogamy
became a fundamental result in the field and the subject of many extensive studies linking the monogamy to no-signalling principle, secure communication and randomness amplification~\cite{Barrett2006, Seevinck2007, Pawlowski2009, Pawlowski2010,  Aolita2012, Silva2015, Cheng2021, Mahato2022}.
However, as shown here, the monogamy principle is not fundamental. Rather, it is a mere consequence of the specific mathematical structure of certain inequalities, which are not universal. 
To support our claim, we develop a systematic method to construct Bell inequalities among $N-1$ observers that do not adhere to the monogamy principle for all $N>3$. Furthermore, we provide an interesting minimalistic polygamous scenario based only on bipartite correlations between all pairs of observers. We show that the simultaneous violation of all inequalities is possible if the number of parties is $N=18$. Recognising the practical challenges associated with generating high-fidelity quantum states in experimental setups, we could still identify inequalities that are violated experimentally by six-qubit Dicke states.
The polygamous nature of quantum nonlocality is therefore proven theoretically and confirmed in experiments.


\section{Three parties and strict monogamy}
\label{sec:CHSH}

\begin{figure}[h]
   \centering
   \includegraphics[width=0.48\textwidth]{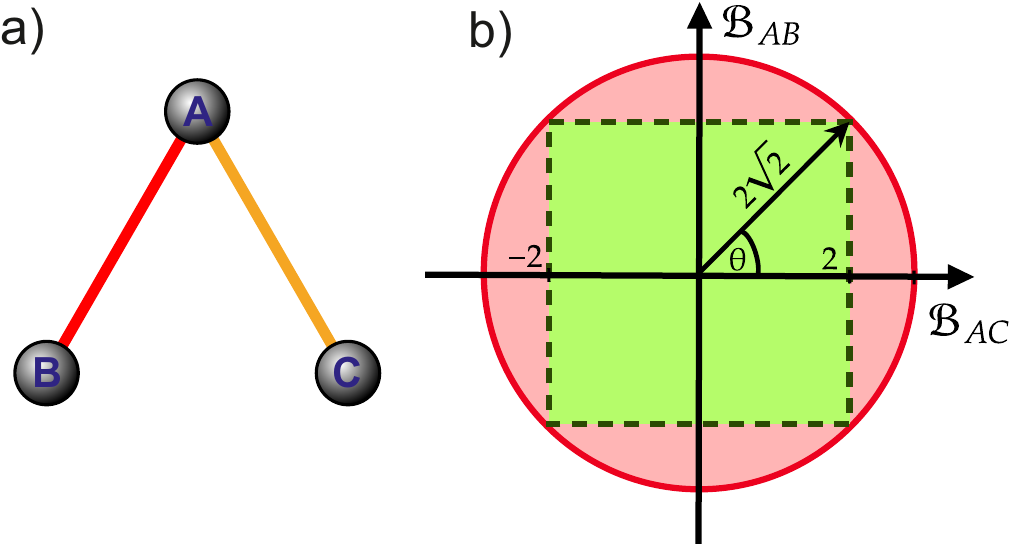}
   \caption{Visual representation of monogamy between CHSH inequality violations for three parties.
   $a)$~Schematic of the arrangement where three observers ($A$, $B$, $C$) try to violate two inequalities (red and orange edges). 
   $b)$  Accessible values of Bell parameters $\mathcal{B}_{AB}$ and $\mathcal{B}_{AC}$ for parties $AB$ and $AC$. According to local realistic models, the bound on each inequality is given by 2 and hence its predictions are confined to the square with a side length of $4$. From~(\ref{MON_A}), quantum predictions lie within a circle of a radius $2\sqrt{2}$. The angle $\theta$ in~(\ref{W1_STATE}) for which both inequalities are simultaneously maximised has been explicitly denoted. The principle of monogamy is clearly satisfied since such a state saturates both local realistic bounds and any attempt to violate one of the CHSH inequalities leads to classically achievable correlations on the other one.}
   \label{fig:circle}
\end{figure}

We begin by recalling the standard results on Bell monogamy between three observers.
Consider a scenario in which party $A$ tries to simultaneously violate the CHSH inequalities \cite{CHSH1969} with parties $B$ and $C$ using a three-qubit quantum state. 
We denote the value of the CHSH-Bell parameters by $\mathcal{B}_{AB}$ and $\mathcal{B}_{AC}$, respectively. Quantum mechanics predicts that these parameters obey the relation~\cite{SG2001,TVmono2006}
\begin{equation}
\mathcal{B}_{AB}^2 + \mathcal{B}_{AC}^2 \le 8.
\label{MON_A}
\end{equation}
Note that if one of the inequalities is violated, e.g. $\mathcal{B}_{AB} > 2$, then the other one cannot be violated, $\mathcal{B}_{AC} < 2$.
This is the statement of monogamy of Bell inequality violations. In the considered scenario, this relation is tight, in the sense that all the values of $\mathcal{B}_{AB}$ and $\mathcal{B}_{AC}$ that reach the bound are realised by quantum theory \cite{TVmono2006,Kurzynski2011}. To see this, consider a pure state given by
\begin{equation}
\ket{\psi} = \frac{1}{\sqrt{2}}(\cos \theta |110\rangle + \sin \theta |101\rangle + |011\rangle),
\label{W1_STATE}
\end{equation}
with $\theta \in [0,\pi/2]$ and the following measurement settings described by the local Bloch vectors
\begin{equation}
\begin{array}{lclcl}
\vec a_1 = \vec x, & \quad & \vec b_1 = \vec c_1 = \frac{1}{\sqrt{2}}(\vec x + \vec y), \\
\vec a_2 = \vec y, & \quad & \vec b_2 = \vec c_2 = \frac{1}{\sqrt{2}}(\vec x - \vec y). 
\end{array}
\end{equation}
Using them, one can evaluate the corresponding two-party Bell parameters as
\begin{eqnarray}
\mathcal{B}_{AB} & \equiv & E_{110} + E_{120} + E_{210} - E_{220} = 2 \sqrt{2} \sin\theta, \nonumber \\
\mathcal{B}_{AC} & \equiv & E_{101} + E_{102} + E_{201} - E_{202} = 2 \sqrt{2} \cos\theta, \nonumber
\end{eqnarray}
where $E_{klm}$ stands for the correlation function (average of the product of local results) and index $0$ indicates the party whose measurement outcomes are not included in computation of the correlations.
This clearly realises all the values in~(\ref{MON_A}) that saturate the bound and gives rise to the circle depicted in Fig~\ref{fig:circle}. The whole circumference is obtained without any permutation of observers and measurement outcomes. However, this is the only scenario where all of the involved inequalities cannot be violated
(see~\cite{Tran2018} for the case of three inequalities).

\section{Four and more parties and polygamy}

\begin{figure}[h!]
   \centering
   \includegraphics[width=0.4\textwidth]{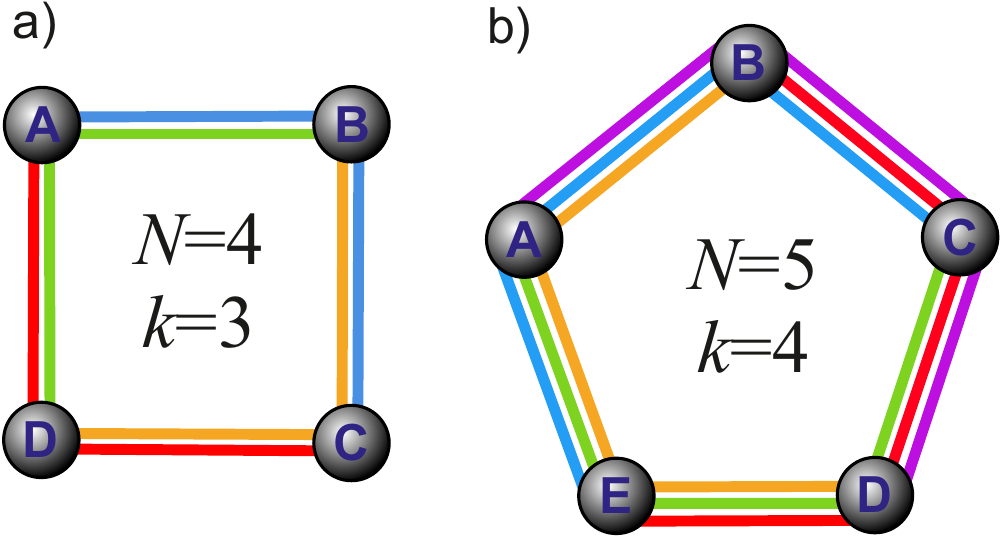}
   \caption{Exemplary configurations showing Bell polygamy. Each colour corresponds to a different choice of $N-1$ observers taking part in the Bell test. Panels $a)$ and $b)$ picture the possible configurations of all three and four-qubit inequalities, respectively. According to our findings, in all such configurations with $N > 3$ one can simultaneously violate all $N$ suitably designed Bell inequalities, each involving $k = N-1$ observers.}
   \label{fig:schemes}
\end{figure}

To demonstrate the polygamous nature of Bell nonlocality, we generalise the Toner-Verstraete scenario~\cite{TVmono2006} to the case of $N$ observers where each of them can perform measurements of two dichotomic observables $A^{(i)}_1, A^{(i)}_2$ ($i=1, \dots, N$).
Then, we analyze the simultaneous violation of Bell inequalities between $N-1$ observers in all possible $N$ configurations (see Fig.~\ref{fig:schemes}). 
Later, we provide an inequality where each of the Bell tests includes measurements between pairs of observers only. 
Following these results, we present an inequality for $N=6$ qubits and demonstrate that using an experimentally observed six-photon Dicke-state one indeed can violate such an inequality for all six five-party subsystems simultaneously.

\subsection{Polygamy of Mermin inequalities}

Our strategy to find Bell inequalities that do not satisfy the monogamy principle is by imposing permutation symmetry. Let $N$ observers share an $N$-qubit quantum state that is permutationally invariant. They perform measurements of the same observables, i.e., $A^{(i)}_j = A^{(1)}_j (j=1,2; i=2,...,N)$. 
In such a situation, if we find a Bell inequality that is violated by a subset of $(N-1)$ observers, it immediately implies that the same inequality is violated by all other subsets with the same number of observers. This does not necessarily exclude correlation trade-offs, but it clearly rules out monogamy.

Let us now use this framework to show that the violation of multi-particle Mermin inequalities is polygamous. To demonstrate this, we consider a general permutation-symmetric pure $N$-qubit state shared by all $N$ observers 
\begin{equation}
|\psi\rangle = \sum_{e=0}^N d_e |D_N^e\rangle,
\label{state}
\end{equation}
where $|D_N^e\rangle$ is the $N$-qubit Dicke state with $e$ excitations.
Now, we form $N$ subsets involving $(N-1)$ parties, and for each of them evaluate the Mermin operator~\cite{MERMIN, Ardehali_1992, Belinskii1993}. For simplicity, justified by symmetry, all further calculations can be performed for observers labelled as $1, 2, ....,  N-1$ only. The Mermin operator can be written as \cite{Belinskii1993, Scarani_2001, Nagata_2006}:
\begin{eqnarray}
M_{N-1} &=& 2^{N-2} \Big(|GHZ^+\rangle_{N-1}\langle GHZ^+| \label{Mermin} \\
    &-&|GHZ^-\rangle_{N-1}\langle GHZ^-| \Big), \nonumber
\end{eqnarray}
where $|GHZ^\pm\rangle_{N-1} = \frac{1}{\sqrt{2}} (|0 \cdots 0\rangle_{N-1} \pm |1 \cdots 1\rangle_{N-1})$. For all local realistic models, the expectation value satisfies $\mathcal{M}_{N-1}=\langle M_{N-1} \rangle \leq 2^{(N-2)/2}$~\cite{Belinskii1993}. Rewriting (\ref{Mermin}) in terms of the Dicke states ($|0...0\rangle_{N-1} = |D_{N-1}^0\rangle$, $|1...1\rangle_{N-1} = |D_{N-1}^{N-1}\rangle$), we get
\begin{eqnarray}
    M_{N-1} &=& 2^{N-2}  (|D_{N-1}^0 \rangle \langle D_{N-1}^{N-1}| \label{MerminDicke} \\ &+& |D_{N-1}^{N-1} \rangle \langle D_{N-1}^{0}|). \nonumber
\end{eqnarray}
In order to calculate the expectation value of the Mermin operator, it is convenient to express the state (\ref{state}) in the following way
\begin{eqnarray}
|\psi\rangle  &=& d_0 |D_{N-1}^0 \rangle |0\rangle \\
 &+& \frac{d_1}{\sqrt{N}} |D_{N-1}^0\rangle |1\rangle 
 + d_1 \sqrt{\frac{N-1}{N}} |D_{N-1}^1\rangle |0\rangle \nonumber\\
 &+& ... \nonumber\\
 &+& \frac{d_{N-1}}{\sqrt{N}} |D_{N-1}^{N-1}\rangle |0\rangle 
 + d_{N-1} \sqrt{\frac{N-1}{N}} |D_{N-1}^{N-2}\rangle |1\rangle \nonumber \\
 &+& d_N |D_{N-1}^{N-1}\rangle | 1 \rangle, \nonumber
\end{eqnarray}
where the dots signify the terms that do not contribute to the expectation value. Finally, we obtain
\begin{eqnarray}
    \mathcal{M}_{N-1} &=& \langle \psi| M_{N-1} \otimes \openone |\psi \rangle \label{EQ_M_STATE}  \\ &=& \frac{2^{N-1}}{\sqrt{N}} (d_0 d_{N-1} + d_1 d_N). \nonumber
\end{eqnarray}
This value is maximal if $d_0=d_{N-1} =0$ and $d_1=d_{N} =1/\sqrt{2}$ or vice versa. Both solutions are locally unitarily equivalent, thus, we choose only the one stated explicitly above. The corresponding state is given as
\begin{equation}
    |\psi_{\max}\rangle = \frac{1}{\sqrt{2}} (|D_N^1\rangle + |1...1\rangle).
    \label{eq:state_max}
\end{equation}
It leads to the maximal quantum value of the Mermin operator equal to 
$\mathcal{M}_{N-1}^{\max}=2^{N-2}/\sqrt{N}$ and violation factor of $2^{(N-2)/2}/\sqrt{N}$. Note that for $N=3$ and $N=4$ our result is consistent with \cite{TVmono2006} and \cite{Kurzynski2011}, and the respective Mermin inequalities are not violated. However, already from $N=5$ onwards such a violation is possible and it increases exponentially with the number of qubits. Remarkably, in the limit of many particles, not only there is no monogamy of violations, but in fact every inequality is violated maximally.

Furthermore, the polygamous property extends to complete sets of tight Bell inequalities, i.e. necessary and sufficient conditions for local realistic models of observed correlations.
We show this for  the WWW\.ZB inequalities~\cite{Werner_2001,Weinfurter_2001,Zukowski_2002}, which are complete for $N$-qubit two-setting correlation functions.
For the proof, consider $(N-1)$-qubit marginal of the state $|\psi_{\max}\rangle$.
Surprisingly, it admits non-vanishing correlations in the $xy$ plane for exactly the same observables as the GHZ state. However, while the GHZ correlations are all perfect, i.e. equal to $\pm 1$, the correlations of the marginal are reduced to $\pm 1/\sqrt{N}$.
It was shown in Ref.~\cite{Zukowski_2002} that such states admit local realistic model if and only if their squared correlations, measured for local observables in a plane of the Bloch sphere,
are larger than $1$. For the marginal state, this sum is equal to $ 2^{N-2}/N$ and is greater than 1 for $N>4$. Note that for $N \le 4$ the correlations do not violate any Bell inequality as they are compatible with the local hidden variable model constructed in~\cite{Zukowski_2002}.

\subsection{Polygamy for four parties}
\label{sec:nonlocpoly4}

As shown in the previous section, it is not possible to violate all four three-qubit Mermin inequalities for observers (ABC, ABD, ACD, BCD) in a four-party system (A,B,C,D), where for simplicity $A^{(1)} \equiv A, A^{(2)} \equiv B$, etc. However, it is possible to find another set of two-setting Bell inequalities that have this feature.

Using an original method based on linear programming (see Appendix~\ref{app}) we found a tight three-qubit inequality $\langle I_{ABC} \rangle \leq 6$, where 
\begin{eqnarray}
I_{ABC}&=&2 \, {\rm sym}[A_1]-{\rm sym}[A_1 B_1] - {\rm sym}[A_1 B_2] \label{eq:ineq_four} \\
&+& {\rm sym}[A_2 B_2] 
 +  2 A_1 B_1 C_1 + {\rm sym}[A_2 B_1 C_1] \nonumber \\
 &-& 2\,{\rm sym}[A_2B_2C_1]  - A_2 B_2 C_2. \nonumber 
\end{eqnarray}
We use here a compact notation for symmetrising over different observers
\begin{eqnarray}
{\rm sym}[A_k B_l C_m] = \sum_{\pi(k,l,m)} A_{k} B_{l} C_{m},
\end{eqnarray}
where the sum is over all permutations of $(k,l,m)$, denoted as $\pi(k,l,m)$, assuming $A_0 = B_0 = C_0 = 1$, e.g., 
${\rm sym}[A_1 B_1]={\rm sym}[A_1 B_1 C_0] = A_1 B_1 C_0 + A_1 B_0 C_1 + A_0 B_1 C_1=A_1 B_1 + A_1 C_1 + B_1 C_1$ being the permutations of $k=1, l=1,m=0$.
Analogous expressions can be formulated for $I_{ABD}$, $I_{ACD}$, $I_{BCD}$ inequalities. This inequality belongs to the {\'S}liwa's set~\cite{Sliwa-2003}.
It can be directly verified that all four subsets of parties can simultaneously violate their inequalities using the four-qubit state of the form $|\psi\rangle = \cos \theta |D_4^1\rangle + \sin \theta |1111\rangle$. Note that this state lies in the same subspace as $|\psi_{\max}\rangle$ in~(\ref{eq:state_max}).
The maximal violation of $6.154 > 6$ is observed for $\theta =  0.144$, and observables lying in the $xz$ plane, $A_i=B_i=C_i=D_i = \cos \phi_i \sigma_x + \sin \phi_i \sigma_z$ with $\phi_1 = 2.739, \phi_2 = 0.847$. It is worth noting that these inequalities can also be simultaneously violated ($6.064>6$) by the Dicke $|D_4^1 \rangle$ state ($\theta=0$) if the angles were chosen as $\phi_1 = 2.640, \phi_2 = 0.986$.
We emphasise the relative simplicity, as every observer measures the same set of two observables.
The above example shows that, already for a system of four particles, one can define inequalities involving three observers such that all of them are simultaneously violated. 

\subsection{Polygamy with two-body correlators}

The violation of monogamy is by no means limited to the case of higher-order correlations. Here we focus on a minimalist scenario based on the measurements of all pairs of observers. Again, we approach this problem through the linear programming technique described in Appendix~\ref{app}. As a result, the $(N-1)$-partite  two-body Bell inequality was found to be $ \langle I_{N-1} \rangle \ge 0$.
The corresponding Bell operator is given as
\begin{eqnarray}
    I_{N-1} &=& L + \alpha \left( {\rm sym}[A_1] + {\rm sym}[A_2] \right) \label{In}  \\ &+& {\rm sym}[A_1B_1] 
 + 2~{\rm sym}[A_1B_2] + {\rm sym}[A_2B_2],  \nonumber
\end{eqnarray}
where $L$ and $\alpha$ are defined by $L = 3((N-4)^2+N-2)$
and $\alpha = -3(N-4)$, respectively.
Note that Eq.~(\ref{In}) fits into the permutationally invariant form of the two-body Bell expressions introduced in \cite{Tura2014}. However, $I_{N-1}$ differs from the two classes of Bell parameters defined in that paper. 

Notably, it is easy to verify the validity of the local realistic bound of zero for (\ref{In}) up to large $N$ (e.g. up to $10^4$). To do this, first note that we can restrict the parties to use deterministic strategies. For a given classical deterministic strategy, let $a_{++}$, $a_{+-}$, $a_{-+}$, and $a_{--}$ denote the number of parties whose classical expectation values $\langle A_k^{(i)}\rangle$ (for $k=1,2$)
are $\{1,1\}$, $\{1,-1\}$, $\{-1,1\}$, and $\{-1,-1\}$, respectively. By definition, we have $a_{++}+a_{+-}+a_{-+}+a_{--}=N$. Due to the symmetry of (\ref{In}), every classical deterministic strategy can be mapped to the four-tuple $(a_{++},a_{+-},a_{-+},a_{--})$. In this way, we reduce the $2^N$-dimensional correlation space to the dimension of 3, and the number of classical deterministic strategies is reduced from $2^{2N}$ to $2(N^2+1)$, which is a significant reduction in the complexity of the problem (for more details, see Tura \emph{et al.}~\cite{Tura2014}). 

To determine the violation of (\ref{In}) one has to minimize the Bell expression in (\ref{In}) for the $N-1$-partite reduced state of some $N$-partite symmetric state, e.g., $\ket{D_{N}^1}$, using observables in the $xz$ plane,
\begin{equation}
A_k = \cos \phi_k \sigma_x + \sin \phi_k \sigma_z,      
\label{Ak}
\end{equation}
where we assume that all observers measure the same pair of observables, i.e. $A_k^{(i)}=A_k$ for all $i$. 

Since $I_{N-1}$ in (\ref{In}) is permutationally invariant, $\ket{D_{N}^1}$ is symmetric, and $A_k^{(i)}=A_k$, the problem of computing the quantum violation is greatly simplified. In fact, we only need to consider the following two-qubit reduced Bell operator
\begin{eqnarray}
    I_{A_1 A_2} &=& L \openone_4 + \alpha (N-1) \left( A_1\otimes\openone_2 + A_2\otimes\openone_2\right)\\
    &+& \binom{N-1}{2}\left(A_1\otimes A_1  + 4A_1\otimes A_2 + A_2\otimes A_2\right),     \nonumber 
\end{eqnarray}
along with the reduced two-qubit state of $\ket{D_{N}^1}$:
\begin{equation}
\rho_2 = \frac{2}{N}\ket{\psi_+}\bra{\psi_+} + \left(1-\frac{2}{N}\right)\ket{00}\bra{00},
\end{equation}
where $\ket{\psi_+}=(1/\sqrt 2)(\ket{01}+\ket{10})$.

Now, by substituting $N=18$ and plugging in the observables defined by the angles
\begin{eqnarray}
\phi_1&=&\pi-\arcsin(21/22), \label{angles} \\
\phi_2&=&\arcsin(21/22).    \nonumber
\end{eqnarray}
we obtain
\begin{equation}
\langle I_{A_1 A_2} \rangle_{\rho_2}= -\frac{4}{99} < 0.    
\end{equation}
Therefore, the two-body Bell inequality (\ref{In}) is clearly violated. This implies that the $18$-qubit state $\ket{D_{18}^1}$ can simultaneously violate all $17$-qubit two-body Bell inequalities and thus the monogamy principle does not hold. Note that here any exchange of information happens only between the pairs of observers.

\section{Experimental demonstration of Bell polygamy}

Although the polygamous character of Bell inequality violations is already present in four-qubit systems, an experimental demonstration of such phenomena would require an experiment with very low experimental errors.
The critical visibility, i.e. the minimal occurence probability of an entangled three-qubit state when mixed with white noise equals $6/6.154 = 0.975$.
For this reason, we construct five-party inequalities for $N=6$ observers with far less demanding visibility requirements to demonstrate experimentally that they are violated in all six five-party subsystems.
Again, by the linear programming method used before we arrive at the tight five-qubit inequality $\langle I_{ABCDE} \rangle \leq 6$ with
\begin{eqnarray}
I_{ABCDE} &=& -~{\rm sym}[A_1B_1] - {\rm sym}[A_2B_2] \label{eq:experimental_ineq} \\&+& {\rm sym}[A_1B_1C_1D_1] + {\rm sym}[A_1B_1C_1D_2] \nonumber \\ &-& {\rm sym}[A_1B_2C_2D_2] + {\rm sym}[A_2B_2C_2D_2]. \nonumber 
\end{eqnarray}
It is worth noting that although these inequalities are defined for five qubits (they constitute the facet of a five-party Bell-Pitovsky polytope), they do not involve five-qubit correlations. 
Interestingly, the reduced five-qubit state of the Dicke state $|D^3_6\rangle$:
 \begin{eqnarray}
    \rho_{\mathrm{nc}} &=& \rm{Tr}_F\left(|D_6^3\rangle_{ABCDEF}\langle D_6^3|\right) \\
    &=& \frac{1}{2}\left(|D_5^2\rangle_{ABCDE}\langle D_5^2| + |D_5^3\rangle_{ABCDE}\langle D_5^3|\right) \nonumber 
    \label{D25} 
\end{eqnarray}
does not contain any five-qubit correlations~\cite{Schwemmer2015,Tran2017}. 
It can be verified that the inequality $I_{ABCDE}$, as well as all of the other inequalities obtained by exchanging the indices, are theoretically violated by the state (\ref{D25}) with a value of $7.8215 > 6$ for the settings $A_i=B_i=C_i=D_i=E_i=F_i=\cos \phi_i \sigma_x + \sin \phi_i \sigma_z$, with $\phi_1 =  1.26042, \phi_2 = 2.77953$. 
The maximal possible violation of~(\ref{eq:experimental_ineq}) by any six-qubit state is comparable with the violation for the Dicke state and is equal to $7.8771$. 
The critical visibility required to observe the violation of inequality (\ref{eq:experimental_ineq}) with the state $\rho_{\mathrm{nc}}$ in Eq.~(\ref{D25}) is $76.71\%$.

We experimentally demonstrate the existence of polygamous Bell-type correlations using five-party subsystems of a six-qubit Dicke state $|D_6^3\rangle$ prepared with polarisation entangled photons.
For this purpose, we use a pulsed laser ($\lambda=390\,\mathrm{nm}$) in an enhancement cavity to exploit the third order of spontaneous parametric down-conversion (SPDC) in a BBO crystal.
By postselecting on six-fold detection events of this collinear type-II SPDC process, we obtain the desired polarization-encoded six-qubit Dicke state with three excitations. Due to the stochastic nature of SPDC the observed events are contaminated by even higher order emissions resulting in a mixture of $|D^3_6 \rangle$ with states of 6 photons traced from $|D^4_8 \rangle$.

\begin{figure*}
\centering
\includegraphics[width=0.85\textwidth]{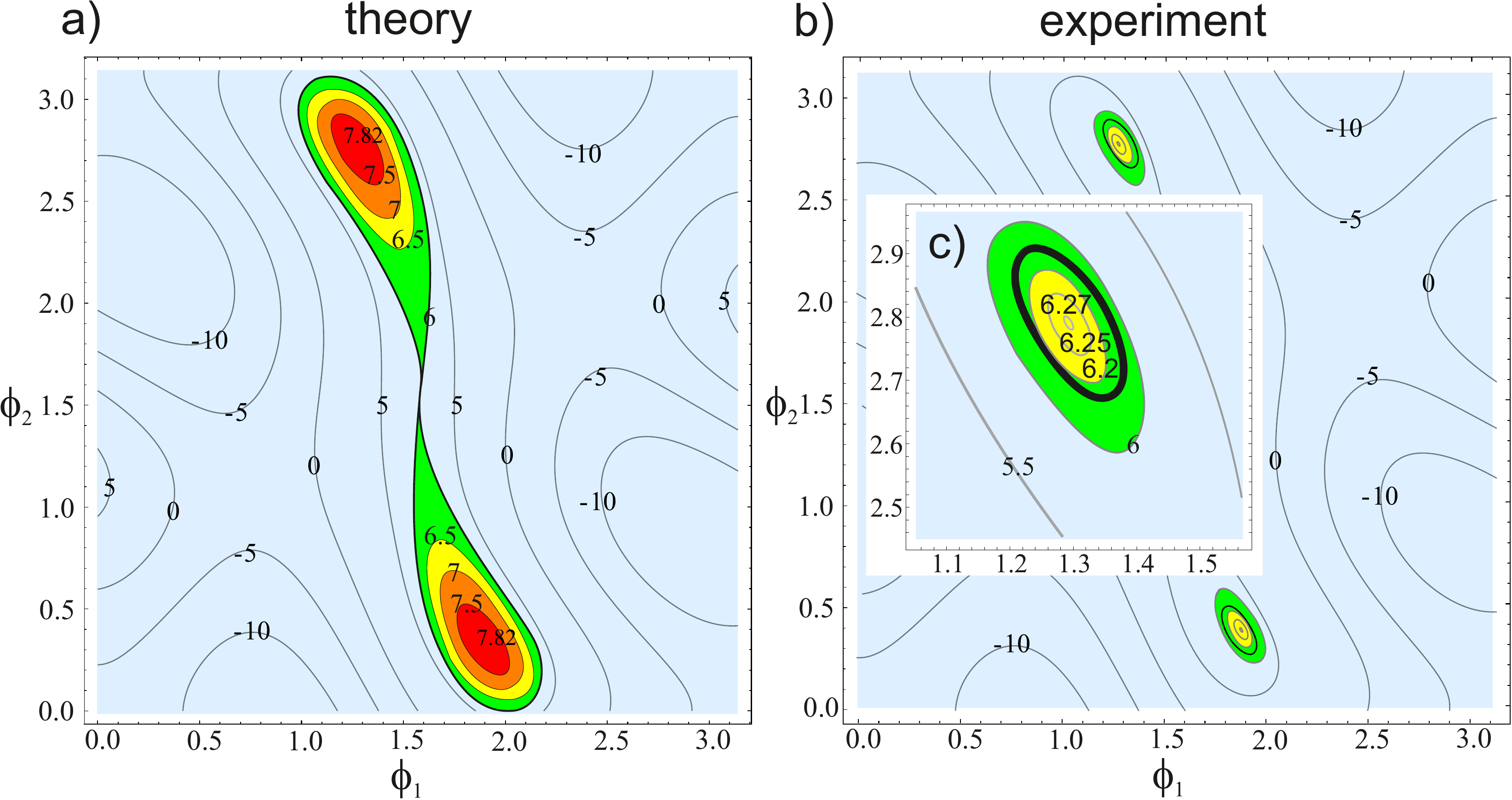}
\caption{\label{figviol} The average violation of inequalities (\ref{eq:experimental_ineq}) plotted as a function of measurement angles $\phi_1$ and $\phi_2$. Panel $a)$ shows the theoretical predictions arising from calculating the violation of inequalities using ideal state $\rho_{\mathrm{nc}}$ in Eq.~(\ref{D25}). Due to the symmetry of the six-qubit Dicke state, the average violation is equivalent to the violation obtained by any set of five parties. The average violation of all inequalities based on the experimental data is presented in panel $b)$. Despite the apparent symmetry, due to the imperfections and statistics, not all inequalities are simultaneously violated in the coloured region. It is the area inside the bold line in panel $c)$ where all six inequalities are violated simultaneously.}
\end{figure*}

\begin{table}[]
\caption{\label{tabviolsim} Experimental polygamy of Bell nonlocality. Simultaneous violation of all six five-party Bell inequalities~(\ref{eq:experimental_ineq}), with the local realistic bound of $6$, where each observer measures the same settings in the $xz$ plane parameterised by $\phi_1 = 1.29244$ and $\phi_2 = 2.79605$. All inequalities are violated by at least one standard deviation.}
\begin{tabular}{c | c c c c c c} \hline
Partition & \footnotesize{ABCDE} & \footnotesize{ABCDF} & \footnotesize{ABCEF} & \footnotesize{ABDEF} & \footnotesize{ACDEF} & \footnotesize{BCDEF} \\ \hline
Violation & 6.315 & 6.204 & 6.479 & 6.307 & 6.146 & 6.128 \\ 
Std. dev. & 0.133 & 0.131 & 0.137 & 0.137 & 0.125 & 0.128 \\ 
\hline
\end{tabular}
\end{table}

A detailed description of the experimental setup can be found in Refs.~\cite{Wieczorek2009,Krischek2010}.
To obtain the density matrices of five-qubit marginals, we evaluate five-fold coincidence events for all subsets of five observers. This provides better measurement statistics compared with tracing out one qubit.
All matrices were obtained using permutation-invariant tomography~\cite{Toth2010}. Due to the different number of detection events and different efficiencies of individual detectors, the five-qubit reduced states are not identical but show small variations.
The average fidelity (averaged over different five-party subsystems) to the ideal $\rho_{\mathrm{nc}}$ state in Eq.~(\ref{D25}) is measured to be $91.2\,\%\pm0.6\,\%$ (between $90.4\,\%\pm0.2\,\%$ and $91.9\,\%\pm0.2\%$). The five-qubit experimental states also show $4.9\,\%\pm0.6\,\%$ average fidelity (between $4.6\,\%\pm0.2\,\%$ and $5.1\,\%\pm0.2\,\%$) to an even mixture of $|D_5^1\rangle$ and $|D_5^4\rangle$ states, originating in the higher-order SPDC emissions. This noise contributes negative values of the Bell parameter in~(\ref{eq:experimental_ineq}) reducing the violations.
Still, all six five-party Bell inequalities are simultaneously violated (see Table~\ref{tabviolsim}) by at least one standard deviation using the same settings parameterised by $\phi_1 = 1.29244$, $\phi_2 = 2.79605$.
The observed average violation is $37.5787/6 > 6$. Figure \ref{figviol} shows how the violation depends on the choice of measurement settings (the same for every observer) both for the theoretical prediction as well as for the experimentally prepared and measured state.
A small, yet clearly visible, region of settings indicates where the simultaneous violation of all inequalities is indeed possible.


\section{Conclusions}\label{sec3}

We showed that quantum nonlocality can be shared among arbitrarily many parties if the number of observers is $N>3$. 
Using an original linear programming method we found several interesting tight inequalities that do not satisfy the monogamy principle. These include polygamous inequalities for $N=4$, inequalities based on two-body correlators for $N=18$ and experimentally verified inequalities for $N=6$, where all six five-party subsets of observers violated the Bell inequality simultaneously.

Our findings demonstrate that Bell inequality violation is an inherently polygamous phenomenon. In fact, only the scenario with three particles obeys monogamy relations. For more observers, any trade-offs between the values of Bell parameters result from a specific mathematical formulation of a given inequality. 
This conclusion has implications on all practical applications linked to Bell inequalities. First of all, security of quantum cryptography can be derived from Bell monogamy~\cite{Pawlowski2010} and it is now evident that the safety becomes inequality dependent. 
Simultaneous violation of all involved inequalities can be exploited for self-testing. To give a concrete example consider the Mermin inequality. One can verify that the violation of a single inequality obtained by the ($N-1$)-qubit marginal of state $| \psi_{\max} \rangle$ derived here can also be obtained by a noisy ($N-1$)-qubit GHZ state. Therefore, the violation of a single inequality cannot be used for self-testing. However, it is easy to check that the noisy GHZ state cannot violate all inequalities. Adding the requirement of simultaneous violation we find that there is only a two-dimensional subspace spanned by the states discussed below (\ref{EQ_M_STATE}) which achieves the maximum. This subspace can therefore be self-tested based on the polygamy of violations. 
Our final example is communication complexity problems. It is well-known that violation of a Bell inequality can be exploited to save on communication of particular distributed-computing problems~\cite{Buhrman_2010}. The possibility of simultaneous violation by any subset of observers translates to the savings in communication for problems where part of the data will be obtained by a subset of computing stations but it is not known in advance which subset will be given the inputs.
We hope these instances will stimulate further work on using the lack of monogamy, especially to test complex quantum networks.

\section*{Acknowledgments}

LK likes to thank Christian Schwemmer for stimulating discussions.
PC is supported by the National Science Centre (NCN, Poland) within the Preludium Bis project (Grant No. 2021/43/O/ST2/02679). 
WL and MK are supported by the National Science Centre (NCN, Poland) within the Chist-ERA project (Grant No. 2023/05/Y/ST2/00005). 
LK and HW acknowledge support by the DFG under Germany’s Excellence Strategy EXC-2111-390814868 and by the BMBF project QuKuK (16KIS1621).
TP acknowledges Xiamen University Malaysia Research Fund (Grant No. XMUMRF/2022-C10/IPHY/0002). TV acknowledges the support of the EU (QuantERA eDICT), and the National Research, Development and Innovation Office NKFIH (Grants No.~K145927, No.~2019-2.1.7-ERA-NET-2020-00003).

\subsection*{Code and data availability}
An exemplary code implementing the technique described in Appendix~\ref{app} and used to find the Bell inequalities presented in the article is available at \url{https://zenodo.org/records/13893953}.
The experimental data that support the findings of this study are available at
\url{https://zenodo.org/records/13685483}. 

\appendix

\section{Linear programming method for constructing Bell inequalities that reveal polygamy} \label{app}

We give a brief description of our method, which is based on linear programming. We illustrate it with permutationally invariant (PI) Bell inequalities with two dichotomic settings per party, and we focus here on three-party inequalities. 
However, this method can be easily generalized to arbitrary Bell inequalities (with any number of settings, parties, and outcomes).

We choose a PI four-qubit pure state $\rho_4=\ket{\psi}\bra{\psi}$. Let us denote its three-qubit reduced density matrix by $\rho_3$. We focus on measurements in the $xz$ plane of the Bloch sphere, 
\begin{equation}
A_k = \cos\phi_k \sigma_x+\sin\phi_k \sigma_z
\label{obs}
\end{equation}
for $k=1,2$. It is also assumed that all three parties perform the same two measurements, that is, $A_k=B_k=C_k$.  Let us define $A_0=B_0=C_0=\openone$. Then we can write a generic two-setting PI Bell operator for three parties as follows:
\begin{equation}
I_{ABC}=\sum_{0\le k\le l\le m\le 2} \alpha_{k,l,m} \, \text{sym}[A_kB_lC_m],
\end{equation}
where as before \text{sym}[X] means the symmetrization of the expression X, and $\alpha_{k,l,m}$ are the Bell coefficients. We identify $-\alpha_{0,0,0}=L$ with the local bound, and then the Bell inequality is given by $\langle I_{ABC}\rangle\le 0$.

Our goal is to find the Bell coefficients $\alpha_{k,l,m}$ of the Bell inequality that gives rise to the largest quantum per local bound $Q/L$, where the symmetrized quantum expectation values 
\begin{equation}
E_Q(k,l,m)=\text{Tr}\left(\rho_3\, \text{sym}[A_k\otimes B_l\otimes C_m]\right) 
\end{equation}
are completely determined by the state $\rho_3$ and the observables $A_k,B_k$ and $C_k$ ($k=1,2$). On the other hand, a local deterministic strategy $\lambda$ is given by a particular choice of $A_k=\pm 1$, $B_k=\pm 1$ and $C_k=\pm 1$, $i=1,2$ (and $A_0=B_0=C_0=1$). We compute the symmetrized products 
\begin{equation}
E_{\lambda}(k,l,m)=\text{sym}[A_kB_lC_m]
\label{Elambdaijk}
\end{equation}
for $0\le k\le l\le m\le 2$, which amount to 10 components. There are $64$ different $\lambda$ strategies, however, due to the symmetry in (\ref{Elambdaijk}) only $\binom{6}{3}=20$ give distinct values in the set of polynomials~(\ref{Elambdaijk}). 

Without loss of generality, we set $\alpha_{0,0,0}=-L=-1$ and solve the following linear programming task to obtain the maximum quantum per local ratio $(Q/L)=q+1$ along with the Bell coefficients $\alpha_{k,l,m}$:
\begin{equation}
q\equiv\text{maximize}
\sum_{0\le k\le l\le m\le 2} \alpha_{k,l,m}E_Q(k,l,m),
\end{equation}
such that
\begin{equation}
\sum_{0\le k\le l\le m\le 2} \alpha_{k,l,m}E_{\lambda}(k,l,m)\le 0,\,\forall\lambda,
\end{equation}
where maximization takes place over $\alpha_{k,l,m}$, $0\le k\le l\le m\le 2$ with the exception of $\alpha_{0,0,0}=-1$.

A solution $q>0$ in the above optimization problem gives a PI Bell inequality with coefficients $\alpha_{k,l,m}$, which can be violated with the reduced three-qubit state $\rho_3$ and observables~(\ref{obs}) specified by the angles $\phi_1$ and $\phi_2$.

\bibliography{sn-bibliography}%

\end{document}